\documentclass[12pt]{article}
\usepackage[numbers,sort&compress]{natbib}
\usepackage{graphicx}
\usepackage{amsmath,graphicx,amssymb,subfigure}
\usepackage{amssymb}
\usepackage{amsmath}
\usepackage{enumerate}
\usepackage{amsfonts}
\usepackage[all]{xy}
\usepackage[section]{placeins}
\usepackage[colorlinks=true,linktocpage=true,linkcolor=blue,citecolor=blue]{hyperref}
\newcommand{\mt}[1]{\textrm{\tiny #1}}
\newcommand{\bea}{\begin{eqnarray}}
\newcommand{\eea}{\end{eqnarray}}
\newcommand{\rh}{r_\mt{H}}
\newcommand{\al}{\tilde{\alpha}}
\begin{document}

\bibliographystyle{hieeetr}

\pagestyle{plain}
\setcounter{page}{1}

\begin{titlepage}

\begin{center}

\vskip 60mm

{\Large {\bf Thermoelectric  DC conductivities with momentum dissipation from higher derivative gravity  }}

\vskip 1 cm

{\large {\bf Long Cheng, Xian-Hui Ge }}\\

\vskip .8cm

{\it Institute of Theoretical Physics and Shanghai Key Laboratory of High Temperature Superconductors, Department of Physics, Shanghai University,
Shanghai 200444, P.R. China}\\
\medskip

\vskip 0.3cm
{\tt  physcheng@shu.edu.cn, \, gexh@shu.edu.cn,}
\vskip .8cm
{\large {\bf Zu-Yao Sun}}\\

\vskip .3cm
{\it  College of Arts and Sciences, Shanghai Maritime University, Shanghai 200135, China.}\\

\medskip

\vskip 0.5cm

{\tt zysun@shmtu.edu.cn }

\vspace{5mm}
\vspace{5mm}

\begin{abstract}

We present a mechanism of momentum relaxation in higher derivative gravity by adding linear scalar fields to the Gauss-Bonnet theory. We analytically computed all of the DC thermoelectric conductivities in this theory by adopting the method given by Donos and Gauntlett in [arXiv:1406.4742]. The results show that the DC electric conductivity is not a monotonic function of the effective impurity parameter $\beta$: in the small $\beta$ limit, the DC conductivity is dominated by the coherent  phase, while for larger $\beta$, pair creation contribution to the conductivity becomes dominant, signaling an incoherent phase. In addition, the DC heat conductivity is found independent of the Gauss-Bonnet coupling constant.
\end{abstract}
\end{center}
 \noindent
\end{titlepage}
\section{Introduction}
The AdS/CFT correspondence  provides a powerful tool in probing many important phenomena of strongly correlated systems in condensed matter physics
\cite{Hartnoll,McGreevy,Herzog:2009}. In
the context of AdS/CFT, many charge transport coefficients such as  DC conductivity, optical conductivity have been computed by considering the near-equilibrium filed theories on
the boundary with gravity dual in the bulk. One can perturb the boundary by a time-dependent field with frequency $\omega$ to obtain the optical conductivity
\cite{Hartnoll,Herzog:2009}. However, under this approach,
when obtaining the DC conductivity with the limit $\omega\rightarrow0$, one will confront the divergence  due to the spatial translation invariance of the homogeneous
gravitational backgrounds involved. Unfortunately, it is well-known that in the real materials, the spatial translation invariance is not preserved i.e. the momentum are not
conserved because of the presence of impurities and lattices.

In order  to extract the finite DC conductivity holographically, many approaches to breaking of the spatial translation invariance in the bulk have been employed.
There are basically two kinds of translational symmetry breaking:  One is to introduce the lattices \cite{Horowitz:201204,Horowitz:201209,Horowitz:2013,Hartnoll:2012,johanna,YiLing:2013,lingprl,Donos:2013eha,YiLing:2014}, part of which relies
on the complicated numerical computation technic in solving PDE,
 or massive term \cite{Vegh,Gouteraux:2014,Davison1,Blake:2013,Blake:20132,WJP:2014,Davison2,Blake:2014yla} or spatial scalar fields \cite{withers57,kim,Davison:2014,Gout¨¦raux:2014,gls} in the
 gravitational background by hand. Another way  is introducing Chern-Simons term \cite{Ooguri} or  pseudo-scalar \cite{Donos:2011} to spontaneously break the translational
 invariance, which will lead to instabilities.

Recently, a new approach to calculation of the DC conductivity has been developed in \cite{Donos:2014uba,Donos1}. This approach does not rely on the zero frequency limit, but
rather than a time-independent electric field as perturbation on the boundary. The DC conductivities can be obtained in terms of the horizon data by analysing regularity
conditions to the holographic model where the momentum dissipation is due to  linear spatial scalar fields. Further discussions on the holographic massive gravity theory and Einstein-Maxwell theory with inhomogeneous, periodic lattices have been studied in \cite{musso} and \cite{Donos2} respectively.

In this paper, we generalize the strategy presented in \cite{Donos:2014uba,Donos1} to calculate the DC conductivities in five-dimensional Einstein-Gauss-Bonnet-Maxwell-linear
scalar field theory with momentum dissipation. The Gauss-Bonnet (GB) term in string effective action appears as the first curvature stringy correction to Einstein-Hilbert action
 when considering the semi-classical  effect, then the higher order terms is dual to the finite corrections to the $1/N$ expansion of field theory on the boundary
 \cite{cai,Mirjam}. So in the framework of AdS/CMT, it is  interesting to investigate the holographic conductivity of the quantum field theories with the higher derivative gravity dual before the string theory is fully understood \cite{cai10,Barclay,JJ,QP}.
  Furthermore, to
 obtain the finite DC conductivities, we will also introduce spatially dependent massless fields which lead to the momentum dissipation \cite{withers57}. Since we will focus on the
 isotropic  bulk metric, we shall include three scalar fields that are linear in all the spatial directions. The anisotropic solutions with one linear axion have been studied in
 \cite{mateos,ch,ch1,tran}, and the discussions for condensed matter with the anisotropic black brane dual can be found in \cite{Donos1,fgw,gls}.

This paper is organised as follows. In section 2, we present the exact solution for Einstein-Gauss-Bonnet-Maxwell gravity with linear scalar fields. Then following
\cite{Donos1}, we calculate the DC electrical conductivity $\sigma$, thermal conductivity $\bar{\kappa}$ and thermoelectric $\alpha$  in terms of horizon data in section 3.
The conclusions are presented in section 4.

\section{Black brane solutions in Einstein-Maxwell-Gauss-Bonnet gravity with linear scalar fields }

We begin with the following five-dimensional action of Einstein-Maxwell-Gauss-Bonnet gravity with three scalar fields.
\bea
S=\frac{1}{2\kappa^2}\int_Md^5x \sqrt{-g}\Big(R-2\Lambda+\tilde{\alpha}\mathcal{L}_{GB}-\frac{1}{2}\sum_{i=1}^{3}(\partial{\phi_i})^2-\frac{1}{4}
F_{\mu\nu}F^{\mu\nu}\Big)\label{action},
\eea
where $2\kappa^2 = 16 \pi G_5$ is the five-dimensional gravitational coupling and $\Lambda=-6$ is the cosmological constant. $\tilde{\alpha}$ is Gauss-Bonnet
coupling constant with dimension $\rm (length)^2 $ and
\bea
\mathcal{L}_{GB}=\left(R_{\mu\nu\rho\sigma}
R^{\mu\nu\rho\sigma}-4R_{\mu\nu}R^{\mu\nu}+R^2\right)
\eea
 is Gauss-Bonnet term \footnote{We follow the conventions of curvatures as in \cite{carroll}}. $\phi_i(x^{\mu}) (i=1,2,3)$ are 3 massless scalar fields and U(1) gauge field strength is defined as $F_{\mu\nu}=(d A)_{\mu\nu} $.

The equations of motion are easily obtained as
\bea
&&\nabla_{\mu}F^{\mu\nu}=0,\nonumber\\
&&\nabla_{\mu}\nabla^{\mu}\phi_{i}=0,\nonumber\\
&&R_{\mu\nu}-\frac{1}{2}g_{\mu\nu}\Big(R+12+\tilde{\alpha}(R^2-4R_{\rho\sigma}R^{\rho\sigma}+R_{\lambda\rho\sigma\tau}R^{\lambda\rho\sigma\tau})\Big)\nonumber\\
&&~~~~~~~~~~~+\tilde{\alpha}\left(2RR_{\mu\nu}-4R_{\mu\rho}R_\nu^{~\rho}-4R_{\mu\rho\nu\sigma}R^{\rho\sigma}+2R_{\mu\rho\sigma\lambda}R_\nu^{~\rho\sigma\lambda}\right)\nonumber\\
&&~~~~~~~~~~~-\sum_{i=1}^{3}\left(\frac{1}{2}\partial_{\mu}{\phi_i}\partial_{\nu}{\phi_i}-\frac{g_{\mu\nu}}{4}(\partial{\phi_i})^2\right)-\frac{1}{2}\left(F_{\mu\lambda}F_\nu^{~\lambda}-\frac{g_{\mu \nu}}{4}F_{\lambda\rho}F^{\lambda\rho}\right)=0\label{eqm}.
\eea
We will consider homogeneous and isotropic charged black brane solutions, and then work with the following planar symmetric ansatz
\bea
ds^2=-f(r)dt^2+\frac{1}{f(r)}dr^2+r^2(dx^2+dy^2+dz^2),\label{metric}
\eea
where the UV boundary is defined as $r\rightarrow\infty$. To obtain the metric homogeneous, we also assume that the scalar fields are  linearly dependent on the three spatial coordinates
\bea
\phi_i(x^\mu)=\beta_{ia}x^a= a_ix+b_iy+c_iz,
\eea
and gauge field as
\bea
A=A_t(r)dt.
\eea
So the Maxwell equations and Einstein equations can be solved exactly
\bea
A_t(r)&=&\mu\left(1-\frac{\rh^2}{r^2}\right),\\
f(r)&=&\frac{r^2}{4\tilde{\alpha}}\left(1-\sqrt{1-8\tilde{\alpha}+\frac{2\beta^2\tilde{\alpha}}{r^2}-\frac{2\beta^2 \al\rh^2}{r^4}+\frac{8\tilde{\alpha}\rh^4}{r^4}+\frac{8\tilde{\alpha}\rh^2\mu^2}{3r^4}-\frac{8\tilde{\alpha}\rh^4\mu^2}{3r^6}}\right),
\eea
where $\mu$ is the chemical potential of the dual field theory on the boundary, $\rh$ is the black brane horizon i.e. $f(\rh)=0$.  The positive constant $\beta^2=\sum_{i=1}^{3}a_i^2=\sum_{i=1}^{3}b_i^2=\sum_{i=1}^{3}c_i^2$ and the constants $\{a_i, b_i, c_i\}$ are satisfy $\sum_{i=1}^{3}a_ib_i=\sum_{i=1}^{3}b_ic_i=\sum_{i=1}^{3}c_ia_i=0$. In terms of the vector notation $(\vec{\beta}_a)_i=\beta_{ia}$ and $\vec{\beta}_a\cdot \vec{\beta}_b=\sum_{i}\beta_{ia}\beta_{ib}$, we have
\bea
\beta^2\equiv \frac{1}{3}\sum_{i=1}^{3} \vec{\beta}_a\cdot \vec{\beta}_a
\eea
 under the condition \bea \label{betacondition}
 \vec{\beta}_a\cdot \vec{\beta}_b=\beta^2 \delta_{ab}.
 \eea

The temperature can be evaluated directly from the Euclidean continuation of the metric (\ref{metric}), that is
\bea
T=\frac{f'(\rh)}{4\pi}=\frac{6\rh^2-\mu^2}{6\pi \rh}-\frac{\beta^2 }{8\pi\rh}.
\eea
Since the entropy of GB black hole satisfies the area formula, from the Bekenstein-Hawking entropy formula, we obtain the entropy density of horizon
\bea
s=\frac{\rh^3}{4G_5}.
\eea

Finally, we discuss the UV and IR behavior of the solution. First, near the UV boundary $r\rightarrow\infty$,
\bea
f(r)\sim\frac{2r^2}{1+\sqrt{1-8\al}}.
\eea
So the effective  asymptotic AdS radial is
\bea
L^2_{\rm eff}=\frac{1+\sqrt{1- 8\al }}{2}
\to  \left\{
\begin{array}{rl}
1   \ , &  \text {for} \ \al \rightarrow 0 \\
\frac{1}{2}  \ , &  \text{ for} \  \al \rightarrow \frac{1}{8}
\end{array}\right.
\,.
\eea
 Note that the Einstein limit is obtained by taking the limit $\tilde{\alpha}\rightarrow0$, in which the
solution (\ref{metric}) reduces to the metric of \cite{withers57}. To understand the geometry near horizon, we define a new coordinate $u$,
\bea
r-\rh=\frac{3\rh^2}{4(3\rh^2+\mu^2)u}.
\eea
At the zero temperature $T=0$, one can readily check that the extremal black brane geometry is topologically equivalent to $AdS_2\times \mathbb{R}^3$:
\bea
ds^2=\frac{L^2}{u^2}(-dt^2+du^2)+\rh^2(dx^2+dy^2+dz^2),
\eea
where $L$ is the curvature radius of $AdS_2$ with
\bea
L\equiv\sqrt{\frac{3\beta^2+4\mu^2}{12(\beta^2+4\mu^2)}}.
\eea
So we can see that in the absence of $U(1)$ gauge field, the extremal black brane geometry can still  be achieved.

Note that for a fixed mass and chemical potential, there is a space of solutions given by the matrices of parameters $\beta_{ia}$, satisfying the constraint (\ref{betacondition}). One can recast the parameter matrix into the
form $\beta_{ia}=\beta \delta_{ia}$ without loss of generality by performing $O(3)$ transformations, corresponding to redefinitions of the coordinates $x^a$ or of the scalars $\phi_i$. In all, the black brane solution is specified by $T$, $\mu$ and $\beta$, since these transformations leave the solutions invariant \cite{withers57}.
\section{DC conductivities}
In this section, we will evaluate the DC electrical conductivity $\sigma$, thermal conductivity $\bar{\kappa}$ and thermoelectric conductivity $\alpha$  in terms of horizon data.

\subsection{Electric conductivity}
In order to compute the conductivities, we consider the perturbations of the form
\bea
&&g_{tx}\rightarrow\delta g_{tx}(r), \nonumber\\
&&g_{rx}\rightarrow r^2\delta g_{rx}(r),\nonumber\\
&&A_x\rightarrow-Et+\delta A_{x}(r),
  \label{pert1}
\eea
and all the other metric and gauge perturbations vanishing. For simplicity of calculation, we set $\beta_{ia}=\beta \delta_{ia}$ as emphasized in the preceding section.  Regarding this fact, it is consistent to set all scalar fluctuations to be vanished except for the one with the linear piece along the direction $x$. We can arbitrarily denote this scalar by $\phi$ and write
\bea
\phi \rightarrow   \phi+\delta\phi(r).
\eea
Then linearizing the Maxwell equation, Einstein equations and Klein-Gordon equation, we can obtain four independent equations of perturbations:
\bea
&&\delta A_{x}''+(\frac{f'}{f}+\frac{1}{r})\delta A_{x}'+\frac{2\rh^2\mu}{fr^3}(\delta g_{tx}'-\frac{2\delta g_{tx}}{r})=0 ,\label{Max1}\\
&&\delta\phi'-\beta\delta g_{rx}-\frac{2E\mu\rh^2}{\beta f r^3}=0, \label{Ein1} \\
&&\delta g_{tx}''+\frac{r^2+4\tilde{\alpha}(f-r f')}{r(r^2-4\tilde{\alpha}f)}\delta g_{tx}'+\frac{8\tilde{\alpha}ff'-r(4f+\beta^2)}{rf(r^2-4\tilde{\alpha}f)}\delta g_{tx}
+\frac{2\rh^2\mu}{r(r^2-4\tilde{\alpha}f)}\delta A_{x}'=0,\label{Ein2}\\
&&\delta\phi''+\frac{3f+rf'}{rf}\delta\phi'-\frac{(3\beta f+\beta r f')\delta g_{rx}}{rf}-\beta\delta g_{rx}'=0,\label{KG}
\eea
where the prime denotes derivatives with respect to $r$. From (\ref{Max1}), one can define a radially conserved current
\bea
J=-\sqrt{-g}F^{rx}=-rf\delta A_{x}'-\frac{2\mu\rh^2}{r^2}\delta g_{tx},\label{current1}
\eea
which is a constant.
The Einstein equation (\ref{Ein1}) simply gives
\bea
\delta g_{rx}=-\frac{2E\mu\rh^2}{\beta^2fr^3}+\frac{ \delta\phi'}{\beta }.\label{grx1}
\eea
Then it is straightforward to see that the equation of motion for $\delta\phi$ can be simplified as
\bea
\frac{(3f+rf')}{rf}\delta \phi'+\delta \phi''=0.\label{KGS}
\eea
To completely determine the solution of perturbation equations, we also need to impose the boundary condition for fluctuations.  Near the UV boundary $r\rightarrow\infty$,
the scalar field perturbation equation (\ref{KGS}) yields the behavior $\delta \phi\sim k_1+ k_2/r^{4}$, we demand the first term vanishes, and
$\delta g_{tx}$ behaves as $  r^{-2}$, which can be seen from (\ref{Ein2}).

Now we consider the asymptotic behavior near the horizon $r=\rh$. Since we  consider the boundary condition at the future horizon, we will use ingoing Eddington-Finklestein
coordinates $(v,r)$ defined as $v=t+\int  \frac{dr}{f(r)}$  here.

First, the gauge field should be regular at the future horizon, which means that $A_x\sim-Ev+...$. So from the (\ref{pert1}), we conclude that $\delta A_{x}$ should satisfy
\bea
\delta A_{x}\sim-\frac{E}{4\pi T}\log(r-\rh)+\mathcal{O}(r-\rh)
\eea
near horizon $r=\rh$. On the other hand,  it is easy to see that the singular part of the metric (\ref{metric}) can be expressed as
\bea
2\delta g_{tx}dvdx-\frac{2\delta g_{tx}}{f(r)}drdx+2r^2\delta g_{rx}drdx
\eea
in the ingoing Eddington-Finklestein coordinates. We can see from (\ref{grx1}) that $ \delta g_{rx}\sim\frac{1}{r-\rh}$ is divergence as $r\rightarrow\rh$. So to obtain the metric non-singular at the horizon, we should require the metric perturbation behaves as
\bea
\delta g_{tx}&\sim& r^2f\delta g_{rx}|_{r\rightarrow\rh}\nonumber \\
&=&-\frac{2E\mu\rh^2}{\beta^2  r}|_{r\rightarrow\rh}+\mathcal{O}(r-\rh).
\eea
Note that we have used the assumption of $\delta \phi$ is regular at the horizon. Since electric current $J$ is radial conserved, the DC electric conductivity can be easily obtained by evaluation of (\ref{current1}) at the horizon:
\bea
\sigma&=&\frac{\partial J}{\partial E}\\ \nonumber
&=&\left(r+\frac{4\mu^2\rh^4}{\beta^2  r^3}\right)\bigg|_{r\rightarrow\rh}\\ \nonumber
&=&\rh+\frac{4\mu^2\rh}{\beta^2}\\ \nonumber
&=&\frac{\pi T}{2}+\frac{2\pi T \mu^2}{\beta^2}+\frac{\sqrt{3\beta^2+6\pi^2T^2+4\mu^2}}{2\sqrt{6}}+\frac{\mu^2\sqrt{6\beta^2+12\pi^2T^2+8\mu^2}}{\sqrt{3}\beta^2}.
\eea

\begin{figure}[htbp]
 \begin{minipage}{1\hsize}
\begin{center}
\includegraphics[height=4cm] {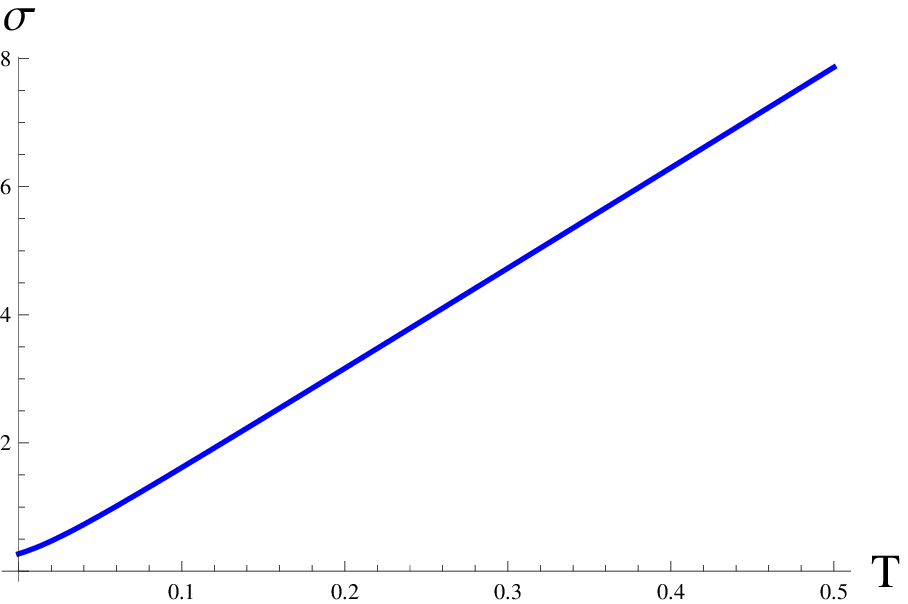}\label{sigmaT}\hspace{6mm}
\includegraphics[height=4cm] {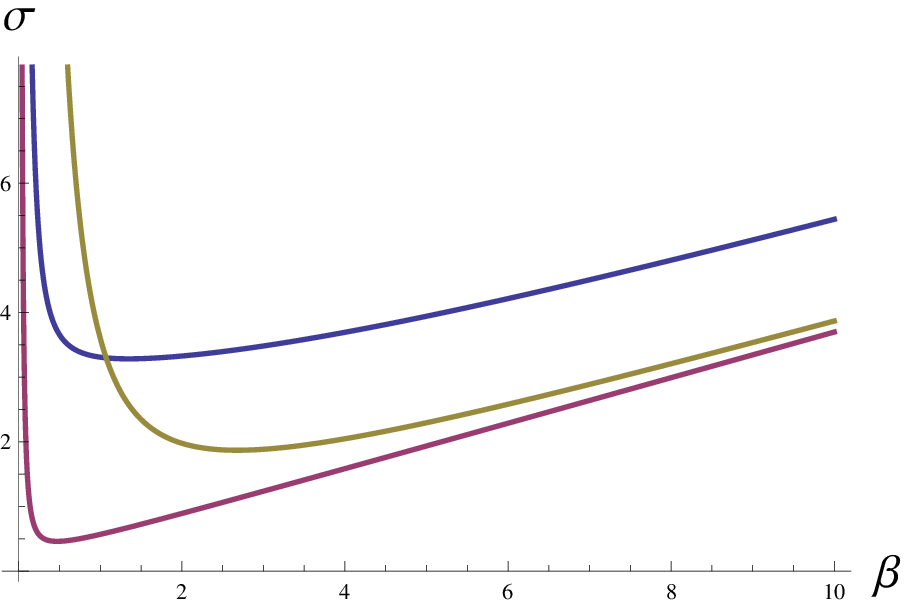}\label{sigmak}\hspace{2mm}
\end{center}
\caption{ On the left is a graph for $\sigma$ as a function of temperature $T$ with $\beta=\mu=0.1$. The right illustrates the scalar parameter $\beta$ dependence
of $\sigma$, where the color lines correspond to $\mu=0.1,T=1$ (blue); $\mu=0.1,T=0.1$ (red) and $\mu=1,T=0.1$ (yellow) respectively.} \label{sigmafig}
\end{minipage}
\end{figure}

As a demonstration, we plot the conductivity as a function of temperature in Fig \ref{sigmafig}. It behaves more like semiconductors, since for
semiconductors, there are insufficient mobile carriers at low temperatures and resistance is high; but as one heats the material, more and more of the lightly bound carriers escape and become free to conduct.
However for normal metals there are plenty of mobile carriers  and the motion of the lattice atoms due to thermal energy causes them to interfere with the transport of mobile carriers through the lattice. Thus, the conductivity of metals decreases as temperature goes up. We can see from Fig.\ref{sigmafig}  that what we obtained does not correspond to normal metals.

The dependence of conductivity on $\beta$ is also shown in Fig.\ref{sigmafig}, in which we can see that in the small $\beta$ limit, $\sigma \propto \frac{1}{\beta^2}$ means that it is dominated by the coherent phase. But as $\beta$ becomes larger, $\sigma \propto \beta$ implies that the contribution of the pair production becomes stronger, leading to an incoherent phase \cite{kim}. This phenomena strongly signals that there is a competition effect between the Drude conductivity and conductivity due to pair creation in the dual field theory.

Now let us examine the behaviour of electric conductivity at low temperature. It is easy to see that, in the limit of $T\ll\mu$, the $\sigma$ behaves as
\bea
\sigma=\frac{(\beta^2+4\mu^2)\sqrt{3\beta^2+4\mu^2}}{2\sqrt{6}\beta^2}+\frac{\pi(\beta^2+4\mu^2)}{2\beta^2}T+...~~~,
\eea
which means that the electric conductivity $\sigma$ is finite as $T\rightarrow0$, indicating the metallic behaviour. On the other hand, for the case $T \gg \mu$, we have
\bea
\sigma=\frac{(\beta^2+4\mu^2)\pi}{\beta^2}T+\frac{3\beta^4+16\beta^2\mu^2+16\mu^4}{24\pi \beta^2}\frac{1}{T}+...~~~.
\eea
To obtain the transport coefficient $\bar{\alpha}$, we should find a conserved heat current analogous to electric current (\ref{current1}). In fact,
we can combine the $tx$-component of Einstein equations and Maxwell equation, and then define a $r$-independent quantity $Q$:
\bea \label{heatc1}
 Q\equiv \frac{\left(r^2-4\al f\right)\left(f\delta g_{tx}'-f' \delta g_{tx}\right)}{r}-A_t J,
 \label{current2}
\eea
which satisfies $\partial_r Q=0$. Different from the Einstein gravity, there is a Gauss-Bonnet coupling constant  contribution in the expression for $Q$.
However, such a $\al$ correction does not contribute to the thermal conductivity as we will see below. As discussed in \cite{Donos1}, the quantity $Q$ should be identical to the heat current in the $x$-direction via calculation of holographic stress tensor, i.e. $Q=T^{tx}-\mu J$. We  present the proof in appendix A where one can see clearly that $Q$ is indeed the heat current.

 Note that $Q$ is independent of $r$, so after evaluation at the horizon, one can obtain $Q=\frac{8E\pi T\rh^2\mu}{\beta^2}$, then $\bar{\alpha}=\frac{\partial Q}{T\partial E}$ is given by
\bea
\bar{\alpha}&=&\frac{8\pi \rh^2\mu}{\beta^2 } \nonumber\\
&=&\pi \mu+\frac{4\pi^3 T^2 \mu}{\beta^2}+\frac{4\pi\mu^3}{3\beta^2}+\frac{2\pi^2T\mu\sqrt{6\beta^2+12\pi^2T^2+8\mu^2}}{\sqrt{3}\beta^2}.
\eea
one may notice that $\al f(r)$ will vanish when evaluating $Q$ on the horizon $\rh$, so $\al$ correction does not contribute to  thermoelectric conductivity.
\subsection{Thermal and thermoelectric conductivities}
To compute the thermoelectric and thermal conductivities, as in \cite{Donos1}, we consider the fluctuations as follows:
\bea
&&g_{tx}\rightarrow t\delta h(r)+\delta g_{tx}(r), \nonumber\\
&&g_{rx}\rightarrow r^2\delta g_{rx}(r),\nonumber\\
&&A_x\rightarrow t\delta a(r)+\delta A_{x}(r),\nonumber\\
&&\phi \rightarrow \phi+\delta\phi (r).  \label{pert2}
\eea
Then, similarly, linearizing the Maxwell equation $\partial_r(\sqrt{-g}F^{rx})=0$, one obtain the following conserved current $\mathcal{J}=-\sqrt{-g}F^{rx}$ with
\bea
\mathcal{J}=-\frac{2\rh^2\mu(\delta g_{tx}+t\delta h )+ r^3 f(t\delta a'+\delta A_x')}{r^2},\label{curr3}
\eea
and the linearised $rx$-component of Einstein equations is given by
\bea
\delta g_{rx}=\frac{2\rh^2\mu \delta a}{\beta^2 r^3f}+\frac{(r^2-4\al f)(r\delta h'-2\delta h)}{\beta^2 r^3f}+\frac{ \delta\phi' }{\beta }.
\eea
Again, to obtain the heat current $\mathcal{Q}$, we also need to know the $tx$-component of the Einstein equations
\bea
&&(r^3f-4\al rf^2)(\delta g_{tx}''+t\delta h'')+\left(4\al f^2+rf(r-4\al f')\right)(\delta g_{tx}'+t\delta h')\nonumber\\
&&+\left(4\al f(rf''-f')+\frac{4\rh^4\mu^2}{r^3}-r^2f'+4\al rf'^2-r^3f''\right)(\delta g_{tx}+t\delta h)+2\rh^2\mu f(t\delta a'+\delta A_x')=0,\nonumber\\
\label{ein3}
\eea
then one can combine (\ref{ein3}) with Maxwell equation, and obtain the conserved current
\bea
\mathcal{Q}=
 \frac{\left(r^2-4\al f\right)\left(f \delta g_{tx}'- \delta g_{tx}f'+  tf\delta h'- t\delta h f'\right)}{r}-A_t \mathcal{J}. \label{curr4}
\eea
In order to calculate the transport coefficients $\alpha$ and $\bar{\kappa}$, we assume $\delta h(r)=-C  f(r) $ and $\delta a(r)=-E+C A_t(r)$ which can be used to cancel the time-dependent terms of the conserved current $\mathcal{J}$ and $\mathcal{Q}$£º
\bea
&&\mathcal{J}\equiv-r(A_t'\delta g_{tx}+f A_x'),\nonumber\\
 &&\mathcal{Q}\equiv \frac{\left(r^2-4\al f\right)\left(f\delta g_{tx}'-f' \delta g_{tx}\right)}{r}-A_t \mathcal{J}.
 \label{current3}
\eea
Similarly, $\mathcal{Q}$ is the time-independent part of the heat current, which will be explained in appendix A.

To find the behaviours of the perturbations near the horizon, we switch to Kruskal coordinates $(U,V)$ instead, which are defined as $U=-e^{-f'(\rh) u/2}$ and $V=e^{f'(\rh)
v/2}$. For the purpose of the metric regularity at the horizon, the perturbation at the horizon should be required as
\bea
&&\delta A_{x}\sim-\frac{E}{4\pi T}\log(r-\rh)+\mathcal{O}(r-\rh),\\
&&\delta g_{tx}\sim  r^2 f \delta g_{rx}|_{r\rightarrow\rh}-C\frac{f}{4\pi T}\log(r-\rh)+\mathcal{O}(r-\rh) \label{gtx1}.
\eea
Note that the positive sign in the first term of (\ref{gtx1}) is chosen to be satisfied the equation for $\delta g_{tx}$.

Now the $\alpha$ and $\bar{\kappa}$ can be easily obtained. First, because $\mathcal{J}$ and $\mathcal{Q}$ are constants in $r$ direction, then evaluating the two conserved currents
(\ref{curr3}) and (\ref{curr4}) at the horizon, we obtain
\bea
&&\mathcal{J}=E\rh+\frac{8C\pi T\mu\rh^2}{\beta^2 }+\frac{4E\mu^2\rh}{\beta^2 },\\
&&\mathcal{Q}=\frac{8\pi ET\mu\rh^2}{\beta^2 }+\frac{16C\pi^2T^2\rh^3}{\beta^2 }.
\eea
Consequently, the conductivities $\alpha$ and $\bar{\kappa}$ are given by
\bea
\alpha &=&\frac{1}{T}\frac{\partial \mathcal{J}}{\partial C}=\frac{8\pi\mu\rh^2}{\beta^2}\nonumber\\
&=& \pi \mu +\frac{4\pi^3 T^2 \mu}{\beta^2 }+\frac{4\pi\mu^3}{3\beta^2 }+\frac{2\pi^2T\mu\sqrt{6\beta^2 +12\pi^2T^2+8\mu^2}}{\sqrt{3}\beta^2 },\\
\bar{\kappa}&=&\frac{1}{T}\frac{\partial \mathcal{Q}}{\partial C}=\frac{16\pi^2 T\rh^3}{\beta^2 }\nonumber\\
&=& 3\pi^3T^2 +\frac{8\pi^5T^4}{\beta^2 }+\frac{4\pi^3T^2\mu^2}{\beta^2 }+\frac{\pi^2T\sqrt{3\beta^2 +6\pi^2T^2+4\mu^2}}{\sqrt{6} }\nonumber\\&&~~~~~~~~~~~~~~+\frac{2\pi^2T(6\pi^2T^2+\mu^2)\sqrt{2\beta^2 +4\pi^2T^2+\frac{8}{3}\mu^2}}{3\beta^2 }.
\eea
The thermal conductivity is the Gauss-Bonnet coupling independent and this   also agrees the previous result obtained in \cite{ge2012}.
At low temperature, these transport coefficients behave as
\bea
&&\alpha=\left(\pi\mu+\frac{4\pi\mu^3}{3\beta^2}\right)+\frac{2\pi^2\mu\sqrt{6\beta^2+8\mu^2}}{\sqrt{3}\beta^2}T+...~~~,\nonumber\\
&&\bar{\kappa}=\frac{\pi^2(3 \beta^2+4\mu^2)^{3/2}}{3\sqrt{6}\beta^2}T+...~~~,
\eea
while at high temperature, the behaviour is
\bea
&&\alpha=\frac{8\pi^3\mu}{\beta^2} T^2+\left(2\pi\mu+\frac{8\pi\mu^3}{3\beta^2}\right)-\frac{\mu(3\beta^2+4\mu^2)^2}{72\pi \beta^2}\frac{1}{T^2}...~~~,\nonumber\\
&&\bar{\kappa}=\frac{16\pi^5}{\beta^2}T^4+\frac{2\pi^3(3\beta^2+4\mu^2)}{\beta^2}T^2+\frac{(3\beta^2+4\mu^2)^3}{864\pi\beta^2}\frac{1}{T^2}+...~~~.
\eea
So we find that thermoelectric conductivity $\alpha$ is finite at $T=0$, while thermal conductivity $\kappa=0$, meaning that a heat gradient does not give rise to transport. Our results imply that we can extend \cite{Donos1} to higher dimensions with higher derivative gravity terms.

It would be interesting to check the Wiedemann-Franz law in our set-up. The Wiedemann-Franz law stated that  the ratio of the electronic contribution of the thermal conductivity  to the electrical conductivity  of a conventional metal, is proportional to the temperature. For this purpose, let us first  introduce the thermal conductivity at zero electric current, which is the usual thermal conductivity that is more readily measurable
$\kappa=\bar{\kappa}-\alpha \bar{\alpha}T/\sigma $ and hence
\begin{equation}
\kappa=\frac{16 \pi ^2 \rh^3 T}{\beta^2+4 \mu ^2}.
\end{equation}
For conventional metals, the Wiedemann-Franz law is characterized by the Lorenz ratio i.e. $L\equiv \kappa/(\sigma T)=\pi^2/3 \times k^2_{B}/e^2$, which reflects that for Fermi liquids the ability of the
 quasiparticles to transport heat is determined by their ability to transport charge so the  Lorenz ratio is a constant.
 Similarly, let us define  the Lorenz ratios as follows
\bea
\bar{L}\equiv \frac{\bar{\kappa}}{\sigma T}=\frac{16 \pi ^2 \rh^2}{\beta^2+4 \mu^2},\\
{L}\equiv \frac{{\kappa}}{\sigma T}=\frac{16 \pi ^2 \beta^2 \rh^2}{\left(\beta^2+4 \mu ^2\right)^2}.
\eea
It is clear that the above equations do not obey the Wiedemann-Franz law  and the Lorenz ratios are not constants.  As $\beta\rightarrow 0$, $\bar{L}$ and $\kappa$ approach finite while $L$ goes to zero and $\bar{\kappa}$ diverges.

\section{Summary}
In this paper, we studied holographic DC thermoelectric conductivities for the higher derivative gravity with momentum relaxation. We presented an exact solution for Gauss-Bonnet-Maxwell theory with
scalar fields. Then we derived analytically the DC electric conductivity, thermal and thermoelectric conductivities of the dual conformal filed on the boundary in the
Gauss-Bonnet-Maxwell theory with momentum dissipation. The exact form of the conductivities confirmed that the approach developed in \cite{Donos1} is applicable even in Gauss-Bonnet gravity in AdS space.

Interestingly, we obtained a Gauss-Bonnet coupling dependent heat currents $Q$ and $\mathcal{Q}$ as be seen in (\ref{heatc1}) and (\ref{curr4}). Unfortunately, such radially independent heat current does not lead to $\al$-dependent thermoelectric and thermal conductivities because the $\al$-dependent term $\al f(r)$ is vanishing at the horizon.

Moreover, different from the conductivities discussed in \cite{kim}, the DC electric conductivity derived in this paper is temperature dependent and basically it  increases as the temperature goes up. The DC electric conductivity does not vanish even at $T \rightarrow 0$ limit.  In our case, at $T=0$ the black brane approaches $AdS_2 \times \mathbb{R}^3$ in the far IR with non-vanishing entropy density. This reflects that the ground states of our system are semiconductors or  bad metals. The electric conductivity at zero temperature might be regarded as arising from charged particle-hole pairs evolution \cite{tong}. This is because in those systems, at higher frequencies,
we can excite electrons from the filled valence band into the conduction band, and these particle-hole pairs then
contribute to the charge density.

It is our interests for the future task to work on the viscosity bound and causality problem in this linear scalar fields modified Gauss-Bonnet theory. The upper bound of the Gauss-Bonnet coupling constant and its relation with the causality has been
investigated in \cite{gb,gb6,gb7,ge1,ge2,ge3}. There are some very recent works on viscosity bound in anisotropic superfluid \cite{roy} and backreaction effects \cite{joshi} in higher derivative gravity. We expect that the presence of the linear scalars may contribute some physics more interesting that would greatly change the
causal structure of the boundary theory and  the upper and lower bounds of the Gauss-Bonnet coupling constant. It would also be interesting to investigate the physics of holographic superconductors in our geometry background by adding a charged scalar field into the action.
\section*{Acknowledgments}
We thank  R. G. Cai, J. X. Lu, Y. Ling,  K.-Y. Kim, S. J. Sin and J. B. Wu for useful discussions at the early stage of this work.
This work was partly supported by NSFC, China (No.11375110).

\appendix
\section{Holographic stress tensor and the heat current}
A standard holographic renormalization procedure \cite{holo1,holo2,holo3} reveals that the holographic stress tensor should be
\bea
\tilde{T}^{\mu\nu}=-2\left(K^{\mu\nu}-K\gamma^{\mu\nu}+3\gamma^{\mu\nu}+2\al(\tau^{\mu\nu}-\frac{1}{3}\tau\gamma^{\mu\nu}) \right) ,
\label{stress}
\eea
 where
\bea
\tau^{\mu\nu}=2 K K^{\mu\lambda}K^{\nu}_{\lambda}-2K^{\mu\lambda}K_{\lambda\rho}K^{\rho\nu}+K^{\mu\nu}(K^{\lambda\rho}K_{\lambda\rho}-K^2),
 \eea
 $\tau$ is the trace of $\tau^{\mu\nu}$ and $K$ is the trace of the extrinsic curvature $K^{\mu\nu}=\nabla^{\mu}n^{\nu}$. Note that we have neglected the term $\frac{1}{4}\gamma^{\mu\nu} \partial \phi_i\cdot \partial\phi_i$ and the Ricci tensor terms which we do not need.

We consider the perturbation (\ref{pert1}) about the black brane, it is straightforward to calculate that
 \bea
\tilde{T}^{tx}&=&\left(\frac{1}{r^2f^{1/2}}-\frac{4\al f^{1/2}}{r^4}\right)\delta g_{tx}'+\frac{2(2f^{1/2}-3r)}{r^3f}\delta g_{tx},
 \eea
and
\bea
\tilde{T}^{xx}=\frac{1}{r^4f^{1/2}}\left(r^2f'-6r^2f^{1/2}+4f(r-\al f')\right),
\eea
where we have used the notation $n_{\mu}=(0,f^{-1/2},0,0,0)$.
So we can deduce that
\bea
f^{1/2}r^3(f\tilde{T}^{tx}-\delta g_{tx}\tilde{T}^{xx})=\frac{(r^2-4\al f)(f\delta g_{tx}-\delta g_{tx}f')}{r}.
\eea
Evaluating both sides at the boundary $r\rightarrow\infty$ and using the expression for $Q$ given in (\ref{current2}), we conclude that
\bea
T^{tx}\equiv r^6\tilde{T}^{tx}=Q+\mu J,
\eea

We now consider the time-dependent perturbation given in (\ref{pert2}) with $\delta h(r)=-C  f(r) $ and $\delta a(r)=-E+C A_t(r)$, then, for the stress tensor of interest, we get
\bea
\tilde{T}^{tx}&=&\left(\frac{1}{r^2f^{1/2}}-\frac{4\al f^{1/2}}{r^4}\right)\delta g_{tx}'+\frac{2(2f^{1/2}-3r)}{r^3f}\delta g_{tx}-Ct\tilde{T}^{xx}\nonumber\\
&\equiv&\tilde{T}_0^{tx}-Ct\tilde{T}^{xx},
\eea
where $\tilde{T}_0^{tx}$ denote the time-independent part of stress tensor. Again, it is easy to check that
\bea
f^{1/2}r^3(f\tilde{T}_0^{tx}-\delta g_{tx}\tilde{T}^{xx})=\frac{(r^2-4\al f)(f\delta g_{tx}-\delta g_{tx}f')}{r},
\eea
and recalling the definition for $\mathcal{Q}$ and evaluating the expression on the boundary $r\rightarrow\infty$, we can conclude that
\bea
T^{tx}-\mu \mathcal{J} =\mathcal{Q}-C t T^{xx}.
\eea

 \end{document}